# Antiproton Stacking in the Recycler

A. Burov, Fermi National Accelerator Laboratory

**Abstract**

Possibilities to accumulate antiprotons in the Recycler are considered for three different cases: with current stochastic cooling, with upgraded stochastic cooling and with electron cooling. With stochastic cooling only, even upgraded, Recycler looks hardly useful. However, with electron cooling at its goal parameters and reasonably good vacuum in the Recycler, this machine would be efficient.

## 1. Introduction

Recycler is a storage ring of 3.3 km constructed to accumulate antiprotons at 8.9 GeV/c, see Ref. [1]. Originally, a significant portion of them (0.2-0.6) were supposed to be antiprotons returned from the Tevatron after their use and significant degradation. That is why the machine got its name. Recently, however, the recycling and 132 ns operation in Tevatron have been dropped from the project scope for a purpose to maximize the performance benefit, while minimizing the cost, effort and technical risk [2]. This revision requests a new scenario for the pbars accumulation. Now, Recycler is supposed to be used for cooling and stacking of pbars coming only from the Accumulator, where the maximal stack is significantly limited by IBS and instability. Both electron and stochastic cooling are supposed to be effectively functioning in the Recycler, providing stacking of a high pbar flux from the Accumulator; this also requires a good lifetime and small diffusion. Presently, the electron cooling is a project with electron beam of required energy being under research and development away from the Recycler, while both longitudinal and transverse stochastic cooling systems are installed and ready to function. That is why the first question of this paper is can the Recycler be useful for the proton accumulation before electron cooling being available; this question is addressed in the next chapter. The second question is about possibilities with the electron cooling, and this is discussed in the chapter after that.

## 2. Stochastic cooling only

Stochastic cooling in the Recycler consists of the longitudinal filter cooling (0.5-1 plus 1-2 GHz) and the transverse cooling (2-4 GHz) systems. A question is how many antiprotons can be accumulated in the Recycler and effectively transferred to Main Injector (MI), assuming certain vacuum, injection / extraction imperfections, IBS, limitations of the Antiproton Accumulator (AA) and MI. Below, this study is described starting from the longitudinal degree of freedom.

## 2.1. Longitudinal Cooling

A process of longitudinal stochastic cooling in a presence of IBS is described by the Fokker-Planck equation (FPE) on the distribution function $F=F(x,t)$ (see, e. g. Refs [3, 4]):

$$\frac{\partial F}{\partial t} = \frac{\partial}{\partial x}\left(\lambda F + \frac{D}{2}\frac{\partial F}{\partial x}\right) + \frac{\partial \phi_{ibs}}{\partial x}$$

with the stochastic cooling force and diffusion given by

$$\lambda = \sum_n \frac{G_n}{\varepsilon_{-n}}, \qquad D = 2\pi N \frac{dx}{d\omega_0}\frac{1}{|n|}\sum_n \left|\frac{G_n}{\varepsilon_{-n}}\right|^2 F$$

where the filters gain $G$ and the beam feedback function $\varepsilon$ are

$$G_n = -i\frac{g_n}{4}\left(1 - \exp\left(-2\pi i n \frac{d\omega_0}{dx}x\right)\right), \qquad \varepsilon_{-n} = 1 - \frac{G_n N}{n}\frac{dx}{d\omega_0}\int dy \frac{F'(y)}{-i(x-y)+0}$$

$$g_n = \begin{cases} g, & \text{for } n_{min} \leq n \leq n_{max} \\ 0, & \text{otherwise} \end{cases}, \qquad n_{max} \leq \frac{1}{2\eta|\Delta p/p|_{max}}, \qquad \eta = \frac{1}{\gamma^2} - \frac{1}{\gamma_t^2} = 0.0088$$

Up to here, $x$ can be anything linearly related to the momentum offset; the distribution $F(x)$ is normalized to $1$. It is convenient to take this variable as a dimensionless phase space area (or longitudinal action) normalized by some "total" phase space as

$$x = A/A_t, \qquad A = 2\frac{\Delta p}{p}\kappa E_0 T_0, \qquad A_t = 30-100 \text{ eVs},$$

where $\kappa = l/C <1$ is the bunching factor (bunch-occupied portion of the orbit for the barrier-bucket bunching). In terms of dimensionless time and the gain,

$$\tau = t/t_{rel}, \qquad 1/t_{rel} = \frac{2(f_{max}^2 - f_{min}^2)\eta A_t}{E_0 N}, \qquad \bar{g} = \frac{\pi^2 gN}{4\omega_0}, \tag{1}$$

FPE rewrites as

$$\frac{\partial F}{\partial \tau} = \frac{\partial}{\partial x}\left(\frac{h_f}{|\varepsilon|^2}\bar{g}xF + \frac{2h_d}{|\varepsilon|^2}\bar{g}^2 x^2 F\frac{\partial F}{\partial x}\right) + \frac{\partial \phi_{ibs}}{\partial x}, \tag{2}$$

where the beam response function

$$\varepsilon = 1 - \frac{2\bar{g}x}{\pi}\int dy \frac{F'(y)}{-i(x-y)+0}$$

lost its dependence on the harmonic number after assumption that the gain is almost linear for core particles, which is an only possibility for the beam coherent response to be significant. The functions

$$h_f = h_f(x) \cong 1, \quad h_d = h_d(x) \cong 1,$$

take into account the gain non-linearity; they are described in the Appendix 1. For a perfectly linear filter and without beam feedback $h_f = h_d = \varepsilon = 1$.

The IBS flux $\phi_{ibs}(x)$ is generally given by the Landau scattering integral, properly averaged over transverse degrees of freedom and over the orbit. This direct approach though would lead to very complicated calculations which never been realized, as we know. Below, instead, an approximation for IBS is suggested which looks both effective and accurate. To start, let it be assumed that instead of the complicated Landau form, IBS flux is described by much more simple Fokker-Planck form:

$$\phi_{ibs}(x) = g_{ibs} x F(x) + \frac{D_{ibs}}{2} \frac{\partial F(x)}{\partial x}$$

A choice of this form is justified by an important feature: it yields a Gaussian distribution as the equilibrium, in agreement with a general theory of scattering. Its parameters $g_{ibs}(x)$ and $D_{ibs}$ follow from Bjorken-Mtingwa (BM) results for emittance growth. Indeed, the emittance growth can be presented as

$$\frac{d\langle x^2 \rangle}{d\tau} = D_0 Q(\mathbf{v}_\parallel).$$

Here $D_0$ gives the emittance growth at zero rms velocity in the beam frame $\mathbf{v}_\parallel = \sqrt{\langle (\Delta p/p)^2 \rangle} \propto \sqrt{\langle x^2 \rangle}$, and $Q(\mathbf{v}_\parallel)$ is a factor describing dependence on this velocity, $Q(0) \equiv 1$. It may be expected that at thermal equilibrium, where longitudinal and transverse rms velocities are equal, $\mathbf{v}_\parallel = \mathbf{v}_\perp$, factor $Q$ goes to 0: $Q(\mathbf{v}_\perp) \cong 0$. Direct calculations with BM theory show that at $\mathbf{v}_\parallel \leq \mathbf{v}_\perp$ this factor can be fitted with very good accuracy as

$$Q(\mathbf{v}_\parallel) = 1 - \sqrt{\mathbf{v}_\parallel / \mathbf{v}_\perp}.$$

This can be seen in Fig. 1, where both direct BM calculations and the fit are presented.

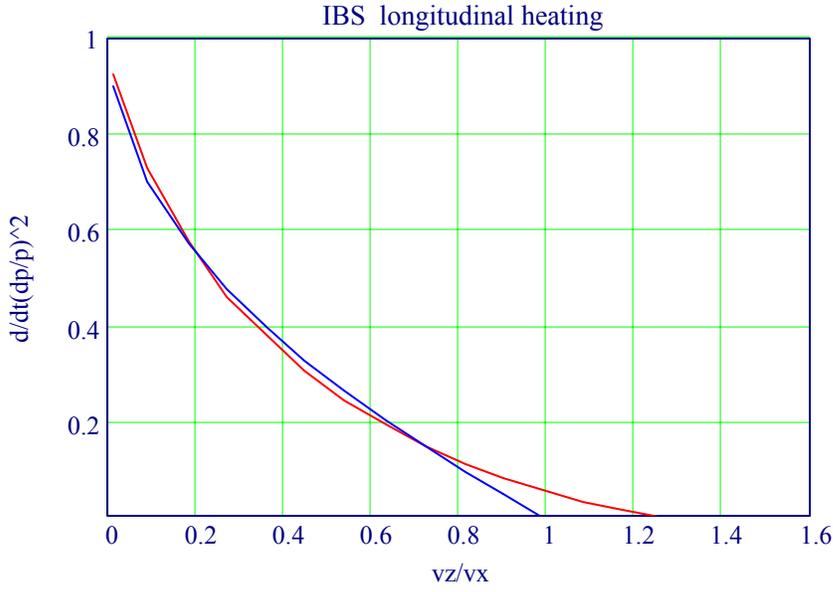

Fig. 1: Longitudinal heating factor $Q$ as a function of the rms velocity ratio $v_\parallel/v_\perp$. Red line is the direct BM results, blue line is the fit.

The parameter $D_0$ gives the IBS diffusion at zero longitudinal temperature:

$$D_0 = \frac{4t_{rel}E_0^2 T_0^2 \kappa^2}{A_t^2}\frac{dv_\parallel^2}{dt} = \frac{2NE_0^3 T_0^2 \kappa^2}{\eta(f_{max}^2 - f_{min}^2)A_t^3}\frac{dv_\parallel^2}{dt},$$

taken at $v_\parallel = 0$. From other side, the emittance growth follows from the IBS flux $\phi_{ibs}(x)$:

$$\frac{d\langle x^2 \rangle}{d\tau} = -2g_{ibs}\langle x^2 \rangle + D_{ibs}$$

A requirement for the two expressions of the emittance growth to be identical leads to

$$D_{ibs} = D_0, \quad g_{ibs} = \frac{D_0}{2\langle x^2 \rangle}\sqrt{\frac{v_\parallel}{v_\perp}}$$

Using that the smooth approximation works fine for Recycler (accuracy 5-8%), the expression for zero-temperature diffusion follows (Ref [5]):

$$\frac{dv_\parallel^2}{dt} = \frac{\sqrt{\pi}Nr_p^2 L_C c v_\perp}{2\kappa C_0 \gamma^2 \varepsilon_n^2}$$

with $r_p$ as the classical proton radius, $L_C \cong 20$ as the Coulomb logarithm, $C_0$ as the ring circumference, and $\varepsilon_n$ as the normalized rms emittance. All the above constitutes a complete definition of the IBS flux in the stochastic cooling plus IBS Fokker-Planck Eq. (2).

More detailed calculations show that the stochastic cooling relaxation time is $\cong 3t_{rel}$ (Eq.1). This parameter does not depend of the bunching factor $\kappa$, being determined by the phase space density. For $A_t \equiv 2A_{rms} = 100$ eVs, $N = 3 \cdot 10^{12}$ and assuming an effective bandwidth as a one-half of the declared with the same central frequency, the relaxation time is $3t_{rel} = 7$ hours. IBS relaxation time can be calculated as ~2-3 times faster at these parameters, assuming required emittances $\varepsilon_n = 1.7 \cdot 10^{-4}$ cm. With these conditions, the stacking is only possible at thermal equilibrium, where all the three (averaged) temperatures are equal, $v_{\parallel} = v_{\perp}$. To provide this condition at given longitudinal and transverse emittances, the bunch has to be squeezed in longitudinal direction to the bunching factor $\kappa = 0.35$. In this case, IBS is mainly reduced to keeping the equilibrium and shaping the distribution; the total 6D emittance growth due to the strong focusing is so slow that can be neglected, ~**50** hours per degree of freedom. Evolution of the distribution function after the last batch with $A = 30$ eVs, $N = 0.75 \cdot 10^{12}$ has been injected from the Accumulator is presented in Fig. 2.

Boundary conditions for the FPE are determined by a finite depth of the barrier bucket potential well. Currently, its **2** kV · **0.9** μs sets $|\Delta p / p\|_{max} = 2.0 \cdot 10^{-3}$; the boundary for simulations of Fig. 2 was supposed to be **1.2** times broader in terms of $\Delta p / p|_{max}$, which requires **50%** increase of the barrier voltage. The longitudinal Schottky band overlap limit gives **30%** wider phase space area than that upgraded barrier bucket.

The final distribution (the blue line in Fig. 2) is close to the equilibrium for a given number of particles, so it is almost independent of the ways how it is reached, such as decrease of the batch initial phase space or its possible gated pre-cooling before merger with the accumulated stack, etc.

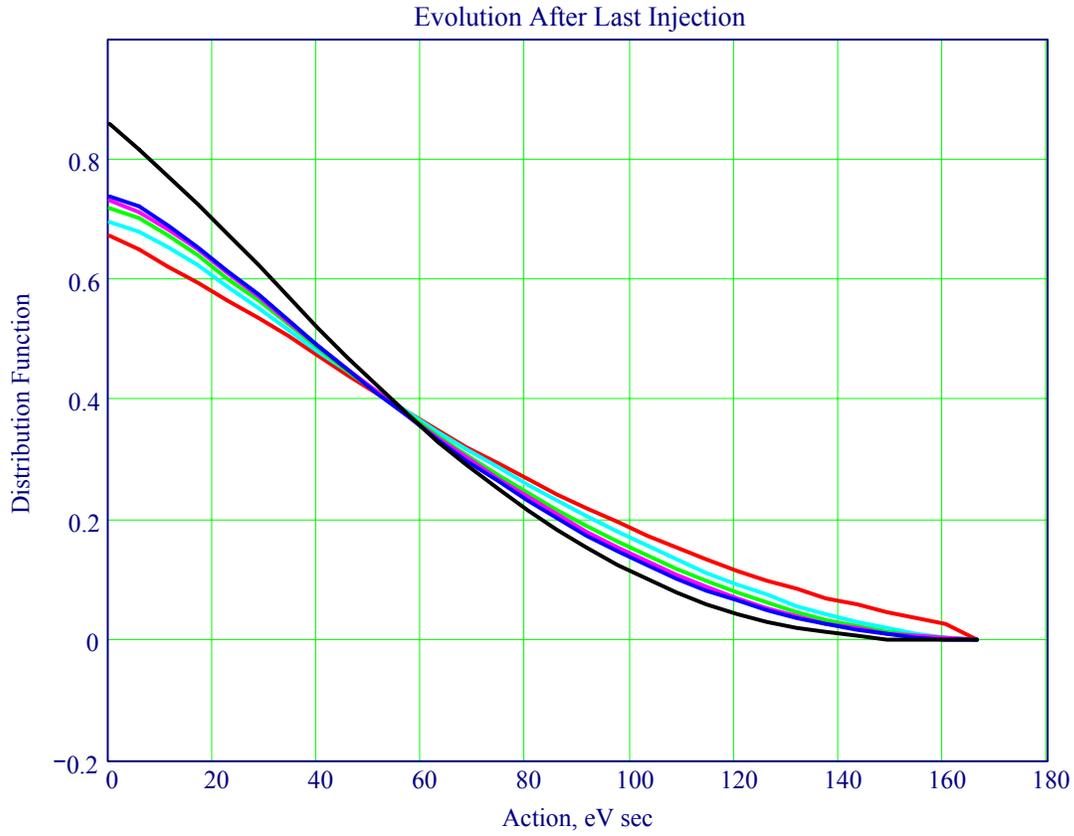

Fig. 2: Evolution at the last stack. Black line shows the distribution before the last injection, the red line is just after injection, and all other lines show how the distribution changes after every ¾ hour. The total final number of particles $N = 3 \cdot 10^{12}$, the time of this process is **3** hours.

## 2.2. Extraction and Longitudinal Losses.

When the final stack is cooled, it is ready for extraction. Beam transfer to the MI is supposed to consist of 9 portions with the bucket capacity of $4 \cdot 3 = 12$ eVs per portion, making $A_{MI} = 12 \cdot 9 = 108$ eVs as the total longitudinal acceptance for the stack in the MI. Efficiency of pbars coalescing in the MI as a function of initial phase space area is presented in Fig. 3.

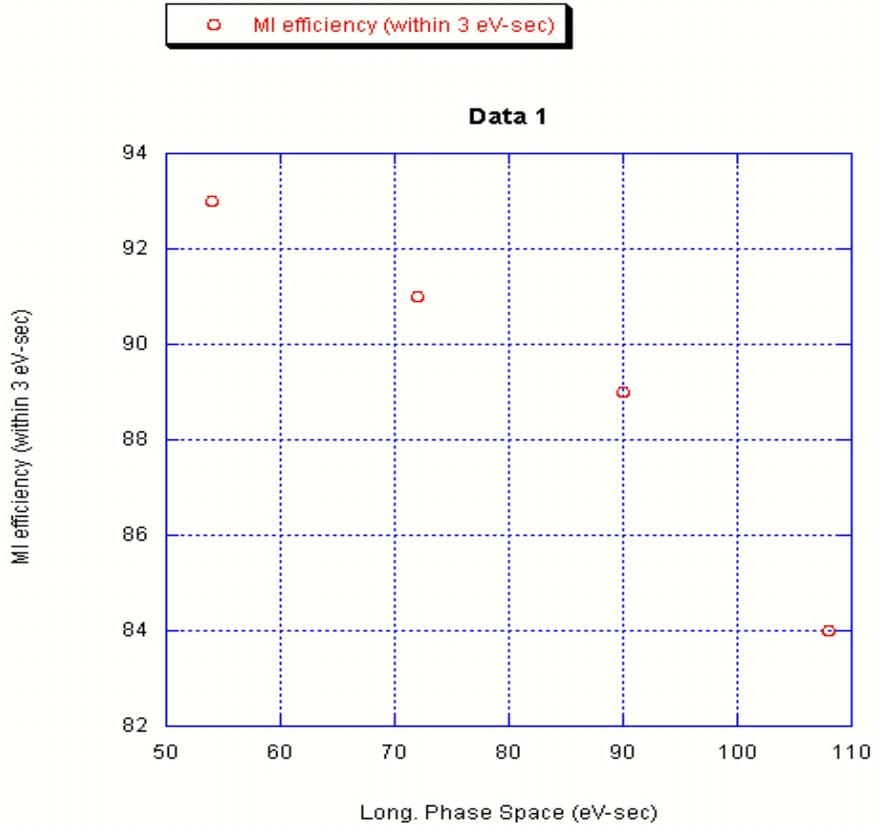

Fig. 3. Efficiency of coalescing as a function of the initial phase space area (by I. Kourbanis).

Fraction of particles outside given phase space for the final beam state (blue line in Fig. 2) is shown in Fig. 4.

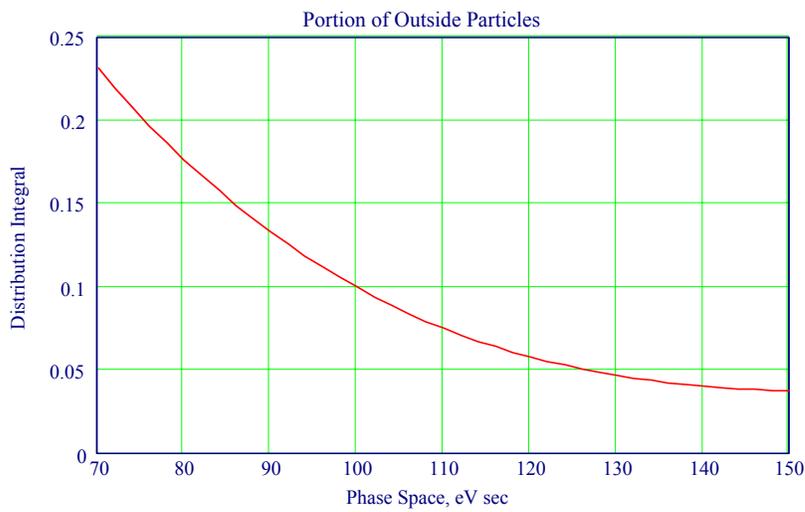

Fig. 4. Fraction of particles outside given phase space area.

Both Figs 3 and 4 give **25%** of the longitudinal reduction of particles after coalescing in MI. This loss figure has to be increased by the transverse finite lifetime losses and finite efficiency of Recycler to MI transfer, which together hardly can be better than 10%. All that means that $3 \cdot 10^{12}$ stacked pbars in the Recycler would give $2 \cdot 10^{12}$ in MI at best. If there is an additional dilution at this extraction as high as **0.5** eVs per every of **36** bunches, the final number of pbars would be $1.7 \cdot 10^{12}$.

## 2.3. Transverse Stochastic Cooling and Losses

Current transverse stochastic cooling in the Recycler has an estimated effective bandwidth from $f_{min} \equiv n_{min} f_0 = 2.5$ GHz to $f_{max} \equiv n_{max} f_0 = 3.5$ GHz. At the optimal gain, beam transverse emittance is cooled with a rate

$$\tau_\perp^{-1} = \frac{2W\kappa}{NM_c}, \quad \text{with} \quad M_c = f_0 F(\Delta f_0 / f_0) \ln(f_{max} / f_{min}) / W,$$

where $W = f_{max} - f_{min}$ is the bandwidth, and $M_c \geq 1$ is the so-called mixing factor expressed in terms of the average normalized longitudinal distribution as a function of the revolution frequency offset, $\int F(\Delta f_0 / f_0) d(\Delta f_0 / f_0) = 1$. Note that the transverse cooling time (as well as the longitudinal) does not depend of the bunching factor, being proportional to the longitudinal phase space density. For $N = 3 \cdot 10^{12}$ pbars inside $A_t = 100$ eVs of the total phase space area, bunched with $\kappa = 0.35$, the mixing factor $M_c \approx 2.0$ leading to the transverse cooling time $\tau_\perp \cong 2.3$ hours. The main transverse heating factor is gas scattering. Were vacuum in the Recycler only two times worse as in the Accumulator, it would have the diffusion $\dot{\varepsilon}_{95\%} = 4$ mm mrad/h leading to $\varepsilon_{95\%} = \dot{\varepsilon}_{95\%} \tau_\perp = 9$ mm mrad of the equilibrium transverse emittance. Thus, having **10** mm mrad for the equilibrium 95% normalized emittance requires for Recycler having that good vacuum. This tough requirement makes the whole scenario with this number of pbars dubious. It would be more realistic to require vacuum not more than 4 times worse than that in the Accumulator ($\dot{\varepsilon}_{95\%} = 8$ mm mrad / h), which would allow accumulation of $N = 1.7 \cdot 10^{12}$ pbars inside $A_t = 100$ eVs and $\varepsilon_{95\%} = 10$ mm mrad.

There is a relation between rms emittance growth $\dot{\varepsilon}$ due to multiple Coulomb scattering and a pencil beam lifetime due to single Coulomb scattering:

$$\dot{\varepsilon} \tau_s = L_s \frac{\varepsilon_{mx} \varepsilon_{my}}{\varepsilon_{mx} + \varepsilon_{my}}$$

Here $\tau_s$ is the pencil beam (zero emittance) single scattering lifetime, $\varepsilon_{mx}, \varepsilon_{my}$ are the ring acceptances and $L_s \approx \ln(10^{-8} / 10^{-13}) \approx 10$ is the scattering Coulomb logarithm.

This relation does not include any gas properties, and can be effectively used to see whether elastic gas scattering is a dominant source of particle losses. For as good vacuum as in the Accumulator, the pencil beam lifetime in Recycler comes out as $\tau_s = 700$ hours. When the beam emittance is not so small, the lifetime is reduced due to multiple scattering, getting more and more important with the emittance over acceptance ratio growth. This dependence of the lifetime on the beam emittance is shown in Fig. 5.

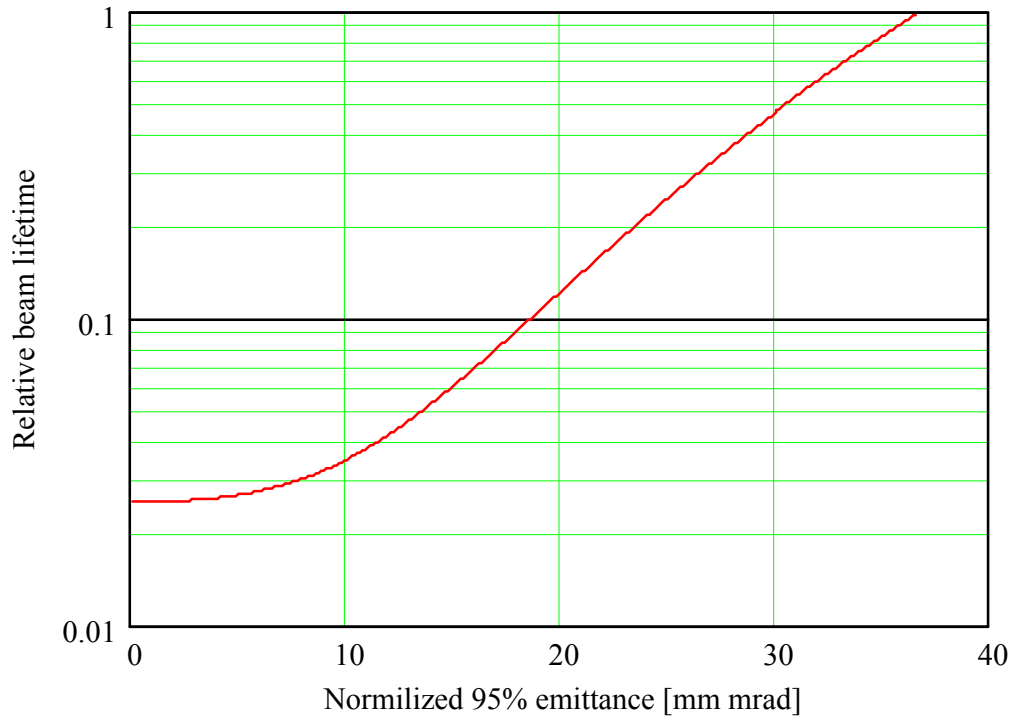

Fig. 5. Relative loss rate as a function of beam emittance.

From this figure, it follows that the lifetime for beam with 10 mm mrad of the normalized 95% emittance is 1.4 times shorter than one of the pencil beam. Thus, with the vacuum which is a factor of 4 worse than that in the Accumulator, the scattering lifetime of that 10 mm mrad beam would be 120 hours. With 20 hours of the stacking time, it leads to 8% of the scattering losses. The total transverse losses includes also inelastic nuclear scattering, which are estimated as +1-2%; thus, the total transverse losses are about 10%, assuming there are no other sources for that.

Counting losses as 5% at extraction plus 10% due to the gas plus 5% at the longitudinal tails leads to 20% of the total losses and $N = 1.4 \cdot 10^{12}$ pbars transferred to the Main Injector.

Possibilities to increase the bandwidth of the stochastic cooling system are very limited: the mixing factor is already rather moderate. That is why the number of pbars after coalescing, with this possible upgrade, is limited by $\cong 2 \cdot 10^{12}$ at best.

## 3. Electron and stochastic cooling together

### 3.1. General considerations

Contrary to stochastic cooling (SC), electron cooling (EC) benefits from phase space reduction. That is why the two cooling systems are conventionally assumed to be complimentary: after SC sufficiently shrinks the beam transversely, EC gets to be efficient. Ultimate temperature of EC is set by either gas scattering or by pbar density factors or by finite angles in the e-beam. From a side of pbars, the low temperature limit of EC can be set by either IBS, or a coherent instability, or the space charge tune shift – all the three phenomena getting stronger with the beam cooling. The first potential stopper, IBS, is going to be excluded in the same way as it was in the previous chapter, namely, keeping the beam at thermal equilibrium. Then, the coherent instabilities can be suppressed by broadband feedbacks, including the SC itself at the highest frequency diapason. Thus, if the vacuum is good, e-beam is aligned, and the first two intensity stoppers are excluded, the beam can be cooled down to the maximal space charge tune shift. For several conventional e-coolers, the beam was cooled to as high tune shift as $\Delta \nu = 0.10 - 0.15$ (see e.g. Ref [6]). When angles of electron trajectories are comparable or higher than pbar ones, it reduces the cooling rates, and even may change their signs. This feature of EC can be used to prevent the beam overcooling, where stability or lifetime can be poor.

To make an effective use of EC, a new batch, before being merged with the accumulated stack, can be pre-cooled transversely by the gated SC. This pre-cooling would be too slow if the longitudinal phase density of the batch is too high. From other side, if the longitudinal phase area of the batch is blown up too much, a burden for the consequent longitudinal EC would be too heavy. Thus, there is an optimal longitudinal phase space area of the batch under the transverse stochastic pre-cooling. EC is not significant at this stage, and the e-beam can be switched off for the batch, which might be also beneficial for the electron current serving the main stack.

After that pre-cooling time passed, the batch transverse distributions have to be shrunk enough; at this moment the pre-cooled batch is merged with the main stack, and the new batch is injected from the Accumulator in its place. To exclude IBS as a significant source of the stack emittance growth, the stack has to be squeezed in the longitudinal direction in accordance to its changing longitudinal and (possibly) transverse emittances. After the merger, the stack has the same repetition time to be e-cooled down to the longitudinal phase space it had before the merger. Transverse gated SC is needed for the stack to compensate lack of EC for high-amplitude particles.

## 3.2. Electron cooling rates

Every time an antiproton passes through the electron beam, it gets a tiny kick against their relative velocity. These kicks, averaged over the betatron phases, yield the EC rates. Generally, the three EC rates (x, y, and z) of the cooled particle are functions of all its three amplitudes; they are expressed in terms of multi-dimensional integrals over the electron velocity distribution, the cooler length and the particle betatron phases; some useful approximations of these integrals can be found in Ref. [7]. For simulations, an analytical fit for the EC rates has been used, where the electron angles were modeled as a transverse temperature described by a certain rms angle in the cooling section. Formulas for this fit of EC rates are expressed in terms of elementary and special (Bessel) functions [8]; they can be found in the Appendix 2. The fit inaccuracy is believed to be not worse than 10-20%. Plots illustrating some features of EC rates are presented in Figs. 6 and 7.

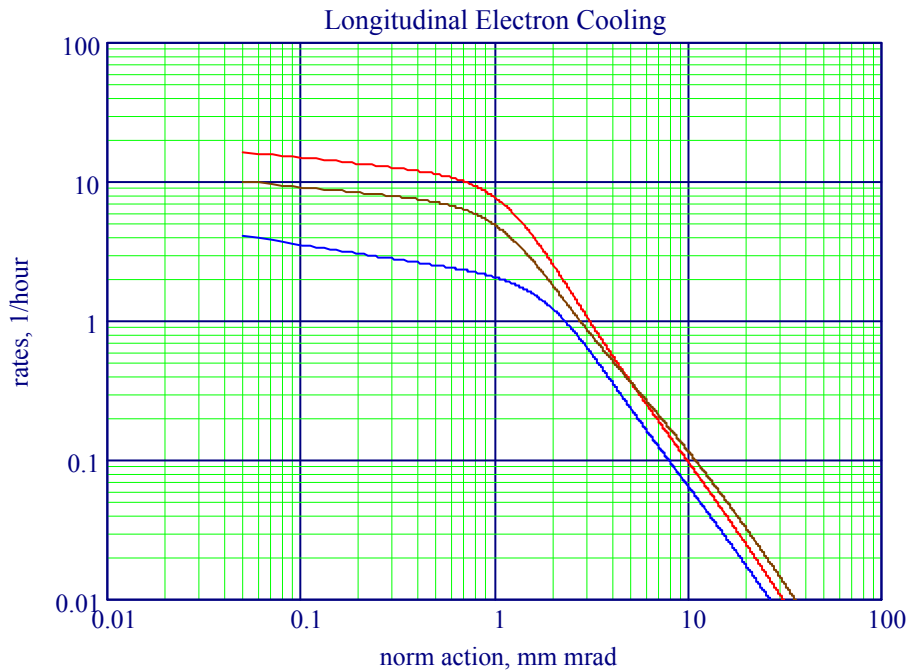

Fig. 6. Longitudinal electron cooling rates as functions of one of the normalized transverse actions, calculated for the cooler parameters of Fig. 1. The transverse action is the Courant-Snyder invariant defined so that its beam average gives the normalized rms emittance. The red line corresponds to the second action of 0.5 mm mrad and with equivalent to that longitudinal velocity in the beam frame. The blue line is for 4 times higher second action and the same longitudinal velocity as the red one. The brown line relates to the same second action and 2 times higher longitudinal velocity compared with the red one.

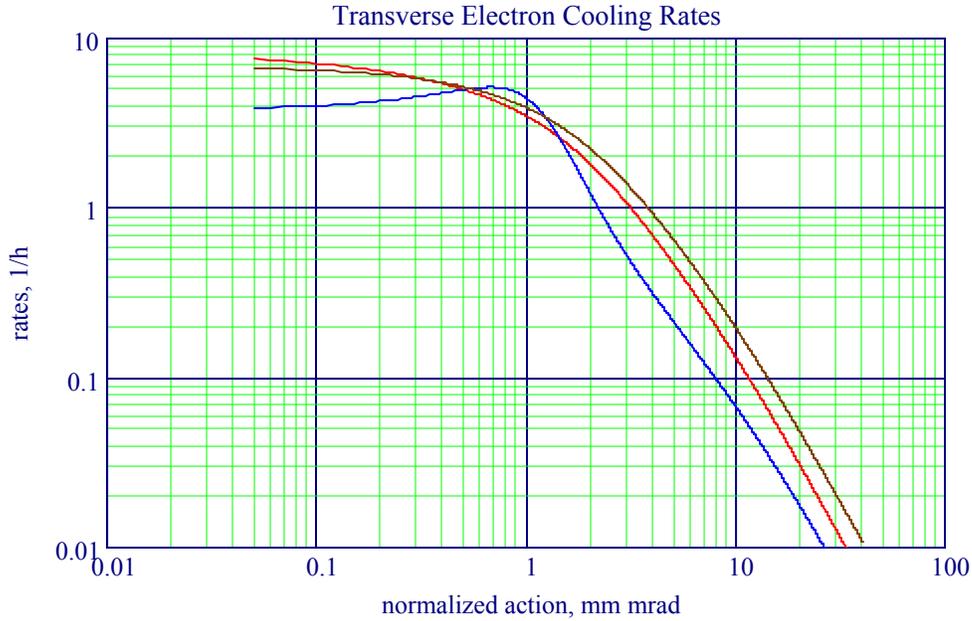

Fig. 7. Transverse EC rates calculated for the same e-cooler parameters as for the previous figure. The red line describes dependence of the horizontal rate on the horizontal action when the vertical action of 0.5 mm mrad and the longitudinal velocity equal to the vertical rms velocity. The blue line shows how the rate changes when the two transverse actions are exchanged with each other. The brown line relates to the same transverse actions as the red one, while the longitudinal velocity is 2 times higher.

## *3.3. Cooling Simulations*

Cooling-stacking process with transverse gated SC and 3D EC of the stack is modeled by Monte-Carlo simulations. The SC with its cooling and diffusion terms renormalized by the feedback through the beam is taken into account in the conventional way, as it is described in the Section 1.3. Electron cooling rates are functions of the three pbar actions, they take into account finite e-beam radius and transverse temperature. The model shows an evolution of the distribution for given values of such input parameters, as initial emittances, transverse and longitudinal diffusion coefficients, injection rate, batch and stack intensities and bunching factors, band of SC and the mentioned e-beam parameters. The simulation consists of two parts: the transverse stochastic pre-cooling of the batch during the repetition period, and then combined electron-stochastic cooling of this batch merged with the stack for the next repetition period. When the stack is merged with the batch, its longitudinal emittance gets to be high. The self-consistency requirement is for the stack emittances being cooled for the repetition period to the values they had just before the last merger. When the final emittances exceed the initial, cooling is insufficient; if they are below, it means that there is an additional safety factor in the cooling. The bunching factor of the pre-cooled batch is not important, because the SC is not sensitive to that, provided the compression is not so high as to drive the bunch into the bad mixing area. Also, the number of particles in the batch is normally considered so small that IBS is not significant for that in any case. As for the stack, the bunching factor is given by the requirement of thermal equilibrium for its current longitudinal and transverse emittances; thus, it varies during the cooling process.

Below, several scenarios of cooling are shown, their common parameters are given in Table 1.

| Transverse stochastic cooling band | **2.5 – 3.5** GHz |
|---|---|
| Batch transverse emittances at injection, 95% norm | **10** $\pi$ mm mrad |
| Repetition time | **1** hour |
| Pbars flux | $\mathbf{45 \cdot 10^{10}}$ /hour |
| Pbars in the stack, up to | $\mathbf{600 \cdot 10^{10}}$ |
| Stack longitudinal 95% phase area | **30** eVs |
| E-cooling length | **20** m |
| Electron 1D rms angle in the cooler | **0.22** mrad |
| Electron beam radius / pbar rms size | **2.5** |
| Beta-function in the e-cooler | **22** m |

Table 1. General parameters of simulations.

*Scenario A1: nominal e-current, nominal vacuum, small emittance.*

Specific parameters for the scenario *A1* are listed in Fig. A1.

| Electron current | **0.5** A |
|---|---|
| Electron beam radius | **2.7** mm |
| Stack 95% normalized emittance | **3** $\pi$ mm mrad |
| Transverse diffusion (norm. 95% emittance growth) | **8** $\pi$ mm mrad /hour |
| Batch 95% longitudinal phase space, inflated to | **60** eVs |

Table A1: parameters of scenario *A1*.

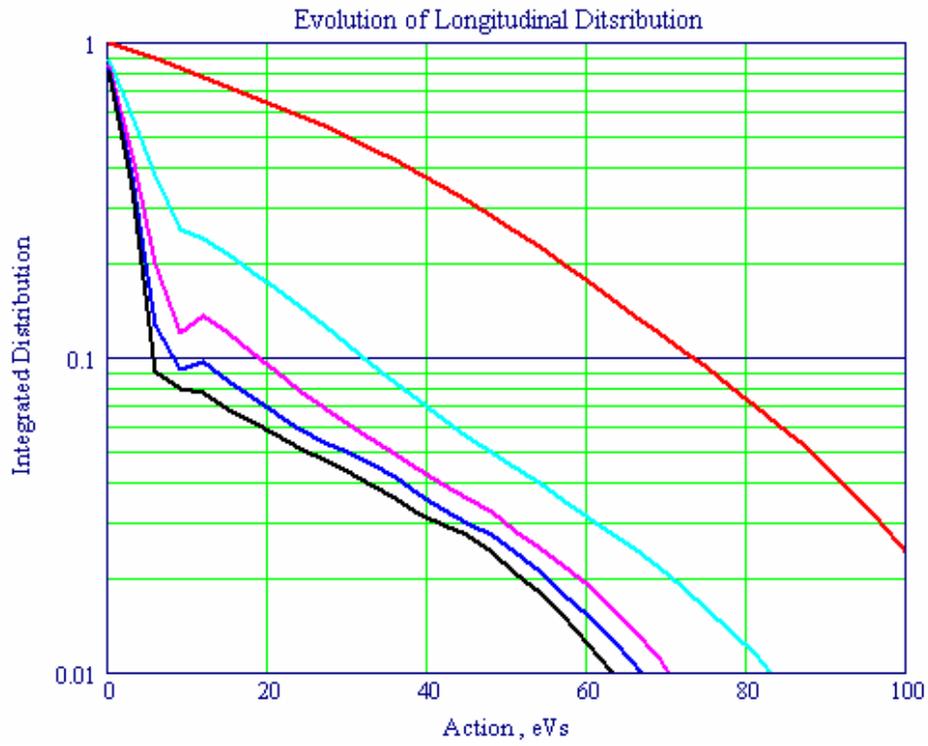

Fig. A1L: Stack plus batch longitudinal evolution is shown for the 60 minutes of EC. The red line shows the state right after the merge, then cyan, magenta, blue and black lines depict the distribution after every 15 min.

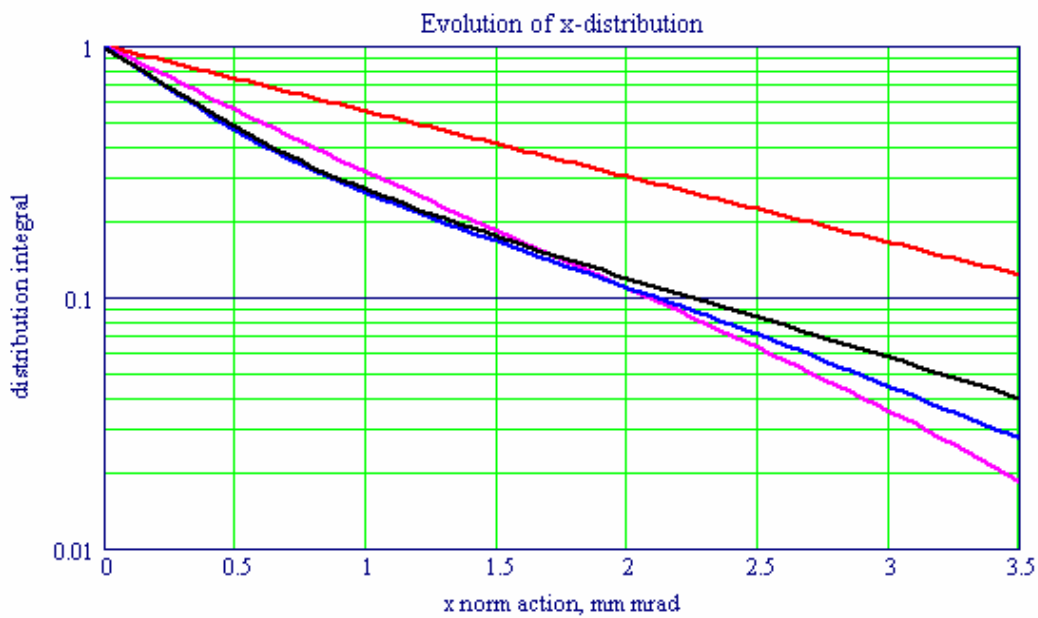

Fig. A1T: The transverse distribution integral (fraction of particles outside a given action) is presented just after injection (the red line), after 1 hour of the gated SC (magenta), after 30 more minutes being merged and cooled with the stack (blue), and right before the next merge (black).

*Scenario A2: lower e-current, higher vacuum, small emittance.*

| Electron current | **0.25** A |
| --- | --- |
| Electron beam radius | **2.7** mm |
| Stack 95% normalized emittance | **3** $\pi$ mm mrad |
| Transverse diffusion (norm. 95% emittance growth) | **5.6** $\pi$ mm mrad /hour |
| Batch 95% longitudinal phase space, inflated to | **60** eVs |

Table A2: parameters of scenario *A2*.

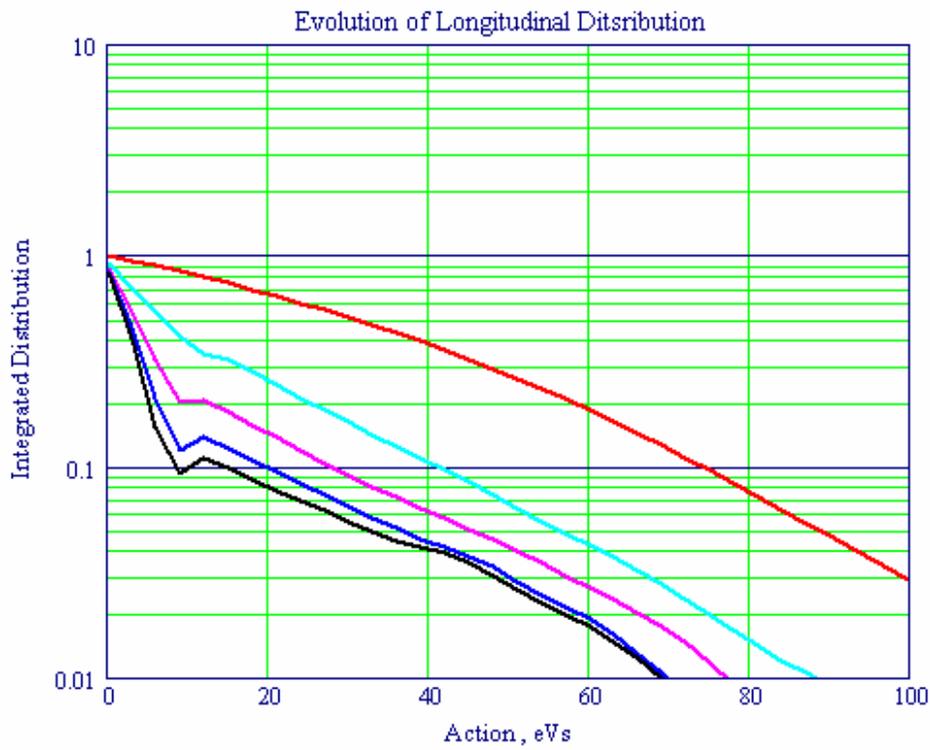

Fig. A2L: longitudinal evolution for A2 scenario: 30 eVs is about total final phase space.

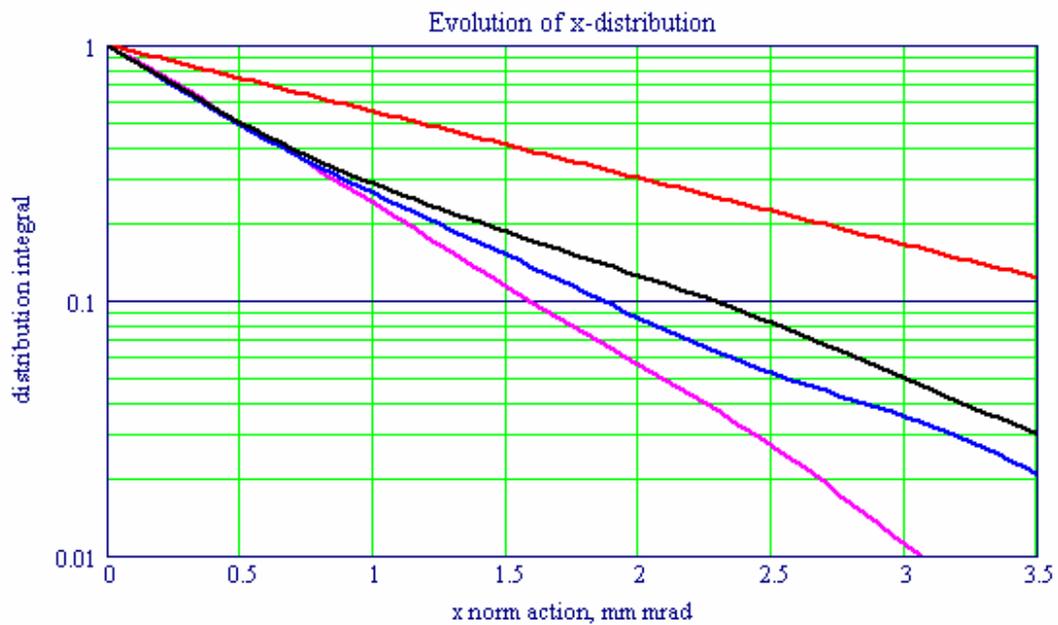

Fig. A2T: Transverse evolution for A2 scenario: beam core is about 1.6 mm mrad of equilibrium rms emittance.

*Scenario B1: nominal e-current, nominal vacuum, nominal emittance.*

| Electron current | **0.5** A |
|---|---|
| Electron beam radius | **5.0** mm |
| Stack 95% normalized emittance | **10** π mm mrad |
| Transverse diffusion (norm. 95% emittance growth) | **8** π mm mrad /hour |
| Batch 95% longitudinal phase space, inflated to | **30** eVs |

Table B1: parameters of scenario *B1*.

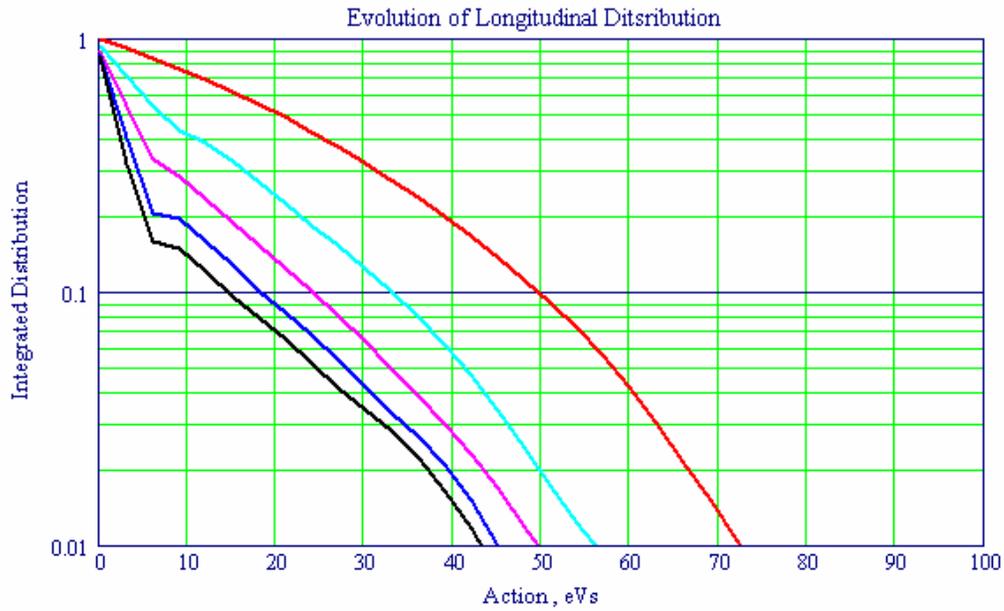

Fig. B1L. Longitudinal evolution for B1 scenario. Final phase space is about 30 eVs, as required.

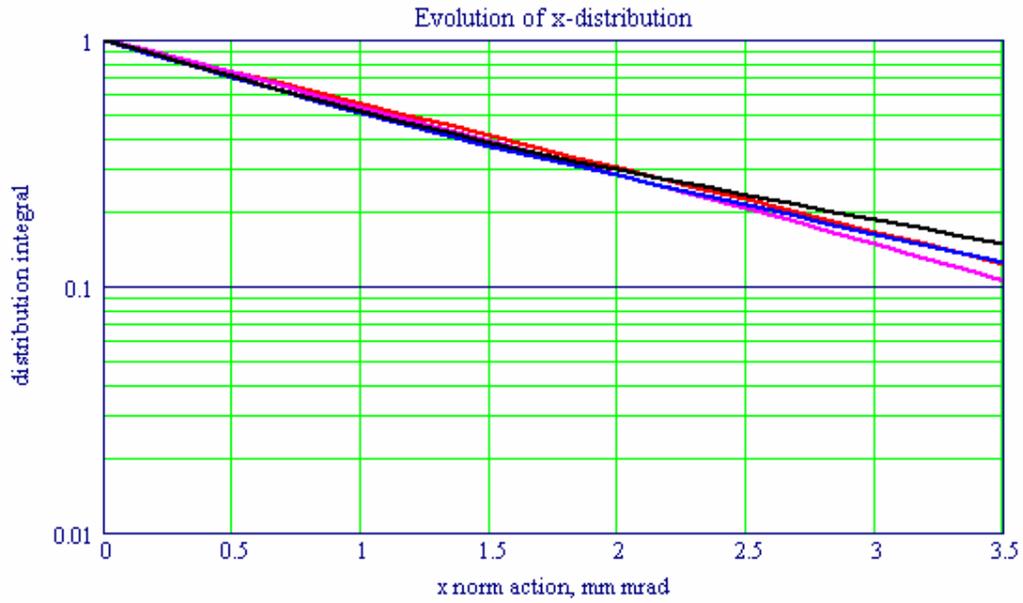

Fig. B1T: Transverse evolution for B1 scenario: no evolution, equilibrium both for SC and EC stages.

*Scenario B2: lower e-current, higher vacuum, nominal emittance.*

| Electron current | **0.25** A |
|---|---|
| Electron beam radius | **5.0** mm |
| Stack 95% normalized emittance | **10** $\pi$ mm mrad |
| Transverse diffusion (norm. 95% emittance growth) | **5.6** $\pi$ mm mrad /hour |
| Batch 95% longitudinal phase space, inflated to | **30** eVs |

Table B2: parameters of scenario *B2*.

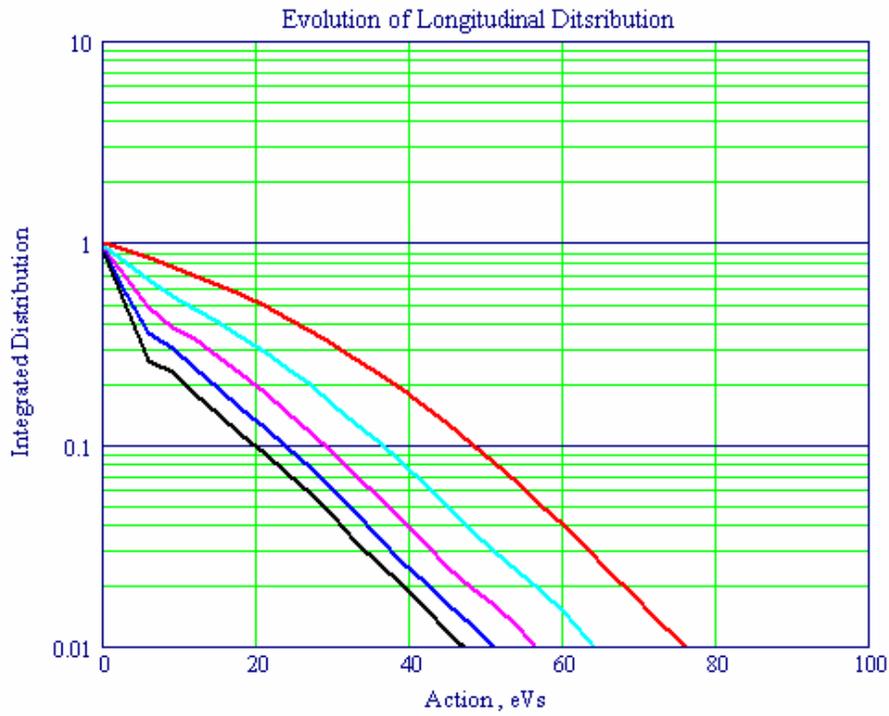

Fig. B2L: longitudinal evolution for B2 scenario.

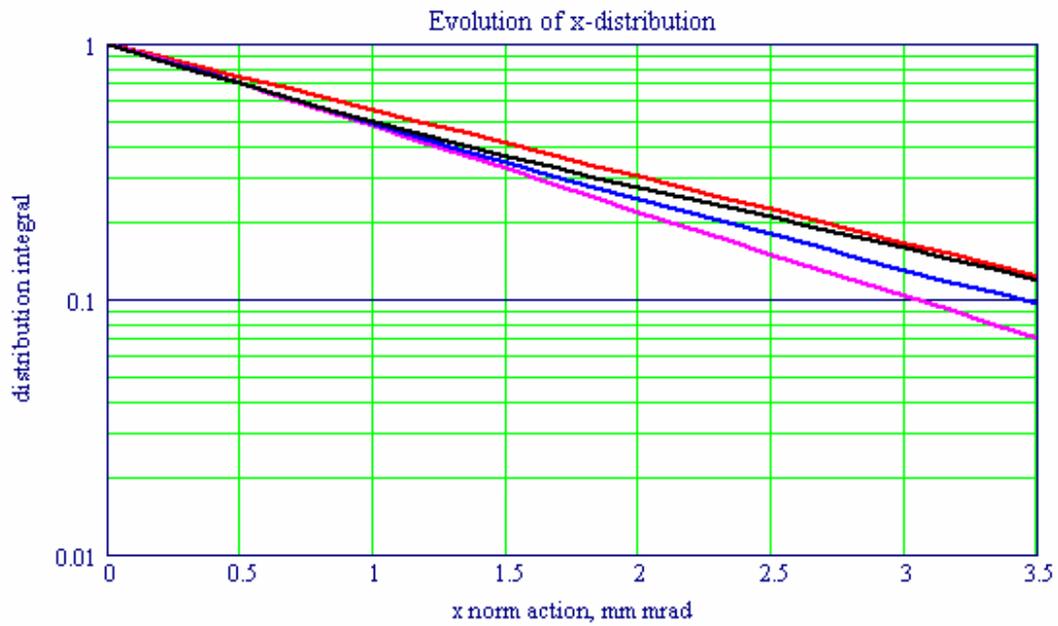

Fig. B2T: transverse evolution for B2 scenario.

Results of this particular simulation present several important features.
- There is infinite number of possibilities to reach the goals of Table 1.

- For the same electron current and vacuum, final stack emittance can be provided as any value between 3 and 10 mm mrad.
- Lower electron current can be compensated by better vacuum
- The stack bunching varies in cooling process. Electron current may be either DC or follow the same pattern.

## *3.4. Coherent Instabilities*

The space charge tune shift $\Delta\nu$ for the maximal number of particles in the cooled stack is calculated as **0.08**, which is not far from its conventional limit of **0.10-0.15**. That high tune shift suppresses Landau damping; thus, the beam is going to be transversely unstable. To prevent this, a broadband feedback is required. The instability, driven by the resistive wall, is expected to be fastest at the lowest frequency, corresponding to the fractional part of the betatron tunes, i. e. at about **50** KHz; the growth time is estimated as **300** turns. The highest limit for unstable frequency band is determined by the Landau damping, being effective at frequencies

$$f \geq f_0 \frac{0.3\Delta\nu}{\eta(\Delta p/p)}.$$

For the listed set of parameters this boundary is as high as **0.7** GHz. Because of uncertainty of the core distribution of the cooled stack, a safety with this issue would require the feedback up to ~ **2** GHz, which is lower frequency of the transverse stochastic cooling system.

## *3.5. Conclusions*

A model is developed which allows simulation of antiproton stacking in the Recycler. It shows that electron cooling might lead to high accumulated current, provided that both the Recycler and the electron beam satisfy certain requirements. Several examples for a set of the satisfactory parameters are presented; details of the cooling process are shown and discussed.

The author is thankful to Dave McGinnis and Valeri Lebedev for essential discussions.

## *3.6. References*

[1] Fermilab Recycler Ring TDR, 1996.
[2] J. Spalding, "Run II Upgrade Program", 2003.
[3] J. Bisognano and C. Leeman, "Stochastic Cooling", in AIP Conf. Proc. No 87, 1982.
[4] D. Möhl, "Stochastic Cooling", in Proc. CAS 1993, CERN 95-06, p.587, 1995.
[5] V. Lebedev et al. NIM-A 391, p.176 (1997).
[6] D. Möhl et al, Proc. ECOOL'99, NIM-A 441, p. 1 (2000).
[7] A. Burov, J. MacLachlan, NIM-A 447, p. 328 (2000).
[8] A. Burov, "Fitting Formulas for Electron Cooling Rates", unpublished note (2003).

## Appendix 1. Special functions for filter-cooling FPE

Beam response function $\varepsilon_{-n}(x)$ is mainly determined by the core particles where the filter is linear. From here, it follows that the response function is almost independent on the harmonic number: $\varepsilon_{-n}(x) = \varepsilon(x)$, and goes out of the harmonic summation in the FPE. The remaining sums depend only on the gain function, but not on the beam distribution, and can be calculated in a general case:

$$\sum_n \text{Re}(G_n) \equiv \frac{n_{max} gy}{4} h_r, \quad \sum_n \text{Im}(G_n) \equiv \frac{n_{max} gy^2}{3} h_i, \quad \sum_n \frac{|G_n|^2}{n} \equiv \frac{g^2 y^2}{16} h_d,$$

where $y = 2\pi n_{max} \eta (\Delta p / p)$ and

$$h_r = \frac{4}{y^2} \sin\left(\frac{f_c}{f_{max}} y\right) \sin\left(\frac{\Delta f}{2 f_{max}} y\right),$$

$$h_i = \frac{6}{y^2} \left[\frac{\Delta f}{f_{max}} - \frac{2}{y} \cos\left(\frac{f_c}{f_{max}} y\right) \sin\left(\frac{\Delta f}{2 f_{max}} y\right)\right],$$

$$h_d \approx \left[1 + (y/4)^{5/2}\right]^{-1},$$

with $f_c = (f_{max} + f_{min})/2$, $\Delta f = f_{max} - f_{min}$.

The special function $h_f$, going with the friction term of the FPE, is expressed as

$$h_f = \text{Re}(\varepsilon(x)) h_r - \text{Im}(\varepsilon) h_i y^2 / 3.$$

## Appendix 2. Electron cooling rates

A set of fitting formulas leading to the EC rates [8] is presented below. The rates are functions of the two transverse actions and momentum offset of the cooled particle, as well as electron beam radius and transverse temperature. Longitudinal temperature of electrons is supposed to be negligible.

### A2.1. Transverse Rates

The transverse cooling rate is defined as $\Lambda_x = -\dfrac{1}{J_x}\dfrac{dJ_x}{dt}$ where $J_x$ is the action in x-direction (the Courant-Snyder parameter, which beam-average is the normalized rms emittance, $\varepsilon_{nx} \equiv \langle J_x \rangle$). The following set of special functions is to be introduced:

$$\mathcal{A}(z) \equiv 2\pi z \exp(-2z) I_0(z)^2$$

$$\mathcal{G}(t) \equiv \dfrac{t}{t^2+1}[1+t+t\ln(t)]$$

$$\mathbb{F}_{in}^{\perp}(x,y,z) \equiv \mathcal{G}\left(\sqrt{\dfrac{x^2+z^2}{y^2+z^2}}\right)(x^2+z^2)^{-3/2}$$

$$\mathbb{F}_{out}^{\perp}(x,y,z) \equiv (2/\pi)(x^2+y^2+z^2)^{-3/2}(x^2+1)^{-1/2}(y^2+1/4)^{-1/2}$$

$$\mathbb{F}_{tot}^{\perp}(x,y,z) \equiv \dfrac{\mathbb{F}_{in}^{\perp}(x,y,z)}{0.7x^{7/2}+3.4y^6-1.8y^{7/2}+1} + \dfrac{\mathbb{F}_{out}^{\perp}(x,y,z)(x^4+y^4)}{x^4+y^4+1},$$

where $I_0(z)$ is the modified Bessel function. After that, the transverse cooling rate can be approximated as

$$\Lambda_x = \Gamma_e \mathbb{F}_{tot}^{\perp}\left(\sqrt{\dfrac{2J_x}{\varepsilon_{ea}}}, \sqrt{\dfrac{2J_y}{\varepsilon_{ea}}}, \dfrac{v_{\parallel}}{v_x}\sqrt{\dfrac{\varepsilon_{nx}}{\varepsilon_{ea}}}\right)\left[\mathcal{A}\left(\dfrac{J_x + \beta_c v_{\parallel}^2/(2\gamma)}{\varepsilon_{eT}}\right)\mathcal{A}\left(\dfrac{J_y + \beta_c v_{\parallel}^2/(2\gamma)}{\varepsilon_{eT}}\right)\right]^{3/4}$$

$$\Gamma_e \equiv \dfrac{4(I_e/e)r_e r_p \eta_c L_{\perp}}{\gamma^{5/2}\varepsilon_{ea}^{5/2}\beta_c^{-1/2}}.$$

Here $I_e$ is electron current, $r_e$, $r_p$ are e- and p- classical radii, $\eta_c = l_c/C$ is a fraction of circumference occupied by e-cooler, $\beta_c$ is the beta-function in the cooler's location, $\gamma$ is the relativistic factor, $\varepsilon_{ea} = \gamma a_e^2/\beta_c$ with $a_e$ as the e-beam radius (of a constant density), $\varepsilon_{eT} = \gamma \theta_e^2 \beta_c$ with $\theta_e \equiv \sqrt{\theta_{ex}^2 + \theta_{ey}^2}$ as the rms angle of electrons in the cooler, $v_{\parallel} = \Delta p/p$ is the pbar longitudinal velocity in the beam frame, $v_x = \sqrt{\varepsilon_{nx}\gamma/\beta_c}$ is its rms transverse velocity (both in units of the speed of light c), and $L_{\perp} = \ln\left(\dfrac{r_{max}}{r_{min}}\right)$ is the Coulomb logarithm for transverse e-cooling (Ref. [7]).

### A2.2. Longitudinal Rates

A sequence of formulae leading to the longitudinal rate $\Lambda_{\parallel} = -\dfrac{1}{v_{\parallel}}\dfrac{dv_{\parallel}}{dt}$ is presented below.

$$F_{in}^{\|}(x,y,z) \equiv \frac{2}{\pi z \sqrt{(x^2 + 2z^2/\pi)(y^2 + 2z^2/\pi)}}$$

$$F_{out}^{\|}(x,y,z) \equiv \frac{1}{\pi xy(x^2 + y^2 + z^2)^{3/2}}$$

$$F_{tot}^{\|}(x,y,z) \equiv \frac{F_{in}^{\|}(x,y,z)}{2x^6 + 2y^6 + 1} + \frac{F_{out}^{\|}(x,y,z)(x^3 + y^3)}{x^3 + y^3 + 1}$$

With these definitions, the longitudinal e-cooling rate is approximated as

$$\Lambda_{\|} = \Gamma_e F_{tot}^{\|}\left(\sqrt{\frac{2J_x}{\varepsilon_{ea}}}, \sqrt{\frac{2J_y}{\varepsilon_{ea}}}, \frac{v_{\|}}{v_x}\sqrt{\frac{\varepsilon_{nx}}{\varepsilon_{ea}}}\right)\left[\mathcal{A}\left(\frac{J_x + \beta_c v_{\|}^2/(2\gamma)}{\varepsilon_{eT}}\right)\mathcal{A}\left(\frac{J_y + \beta_c v_{\|}^2/(2\gamma)}{\varepsilon_{eT}}\right)\right]^{1/2}$$

The presented formulas for the transverse and longitudinal rates have correct asymptotic behavior and agree with the direct integral calculations with accuracy better than 20%.